\documentclass{article}
\usepackage{smc}

\usepackage{times}
\usepackage{ifpdf}
\usepackage{amsmath,cite,url}
\usepackage{color}
\usepackage{bbold}
\usepackage{multirow}
\usepackage[table,xcdraw]{xcolor}
\usepackage[normalem]{ulem}

\makeatletter
\newcommand{\argmin}{\mathop{\text{~argmin~}}}
\newcommand{\argmax}{\mathop{\text{~argmax~}}}

\definecolor{gris}{gray}{0.90}
\definecolor{gris25}{gray}{0.90}
\definecolor{americanrose}{rgb}{1.0, 0.01, 0.24}
\definecolor{bostonuniversityred}{rgb}{0.8, 0.0, 0.0}
\definecolor{shamrockgreen}{rgb}{0.0, 0.62, 0.38}
\definecolor{selectiveyellow}{rgb}{1.0, 0.73, 0.0}
\definecolor{royalblue}{rgb}{0.25, 0.41, 0.88}
\definecolor{ashgrey}{rgb}{0.7, 0.75, 0.71}
\definecolor{burgundy}{RGB}{159,29,53}
\definecolor{darkgreen}{RGB}{18,53,26}
\definecolor{lightblue}{RGB}{102,217,255}
\definecolor{fakeorange}{RGB}{255,140,102}
\definecolor{arylideyellow}{rgb}{0.91, 0.84, 0.42}
\definecolor{bananayellow}{rgb}{1.0, 0.88, 0.21}
\definecolor{gris_f}{gray}{0.35}
\definecolor{bordure}{rgb}{0.09,0.17,0.68}
\definecolor{aquamarine}{rgb}{0.5, 1.0, 0.83}
\definecolor{apricot}{rgb}{0.98, 0.81, 0.69}
\definecolor{babyblue}{rgb}{0.54, 0.81, 0.94}
\definecolor{uipoppy}{RGB}{225, 64, 5}
\definecolor{uipaleblue}{RGB}{96,123,139}
\definecolor{uiblack}{RGB}{0, 0, 0}
\definecolor{decoda}{RGB}{0,153, 0}
\definecolor{lightgreen}{rgb}{0.56, 0.93, 0.56}
\definecolor{blue_f}{rgb}{0.2, 0.2, 0.6}
\definecolor{cinnamon}{rgb}{0.82, 0.41, 0.12}
\definecolor{darkpastelgreen}{rgb}{0.2, 0.75, 0.24}
\definecolor{drab}{rgb}{0.59, 0.44, 0.09}

\def\papertitle{SEMI-SUPERVISED CONVOLUTIVE NMF FOR AUTOMATIC PIANO TRANSCRIPTION}
\def\firstauthor{Haoran Wu}
\def\secondauthor{Axel Marmoret}
\def\thirdauthor{J\'er\'emy E. Cohen}

\newif\ifpdf
\ifx\pdfoutput\relax
\else
   \ifcase\pdfoutput
      \pdffalse
   \else
      \pdftrue
\fi

\ifpdf 
  \usepackage[pdftex,
    pdftitle={\papertitle},
    pdfauthor={\firstauthor, \secondauthor, \thirdauthor},
    bookmarksnumbered, 
    pdfstartview=XYZ 
   ]{hyperref}

  \usepackage[pdftex]{graphicx}
  \DeclareGraphicsExtensions{.pdf,.jpeg,.png}
  \usepackage[figure,table]{hypcap}

\else 
  \usepackage[dvips,
    bookmarksnumbered, 
    pdfstartview=XYZ 
  ]{hyperref}  

  \usepackage[dvips]{epsfig,graphicx}
  \graphicspath{{./figures/}}
  \DeclareGraphicsExtensions{.eps}

  \usepackage[figure,table]{hypcap}
\fi

\hypersetup{
    colorlinks,%
    citecolor=black,%
    filecolor=black,%
    linkcolor=black,%
    urlcolor=black
}

\title{\papertitle}

 \threeauthors
   {\firstauthor} {INSA Rennes, France \\ %
     {\tt \href{mailto:Haoran.Wu@insa-rennes.fr}{haoran.wu@insa-rennes.fr}}}
   {\secondauthor} {Univ Rennes, Inria, CNRS,\\ IRISA, France \\ %
     {\tt \href{mailto:axel.marmoret@inria.fr}{axel.marmoret@inria.fr}}}
   {\thirdauthor} { Univ Lyon, INSA-Lyon, UCBL, \\ UJM-Saint Etienne, CNRS, Inserm, \\
CREATIS UMR5220, U1206, France \\ %
     {\tt \href{mailto:jeremy.cohen@cnrs.fr}{jeremy.cohen@cnrs.fr}}}

\begin{document}
\capstartfalse
\maketitle
\capstarttrue

\sloppy

\begin{abstract}

Automatic Music Transcription, which consists in transforming an audio recording of a musical performance into symbolic format, remains a difficult Music Information Retrieval task. In this work, which focuses on piano transcription, we propose a semi-supervised approach using low-rank matrix factorization techniques, in particular Convolutive Nonnegative Matrix Factorization. In the semi-supervised setting, only a single recording of each individual notes is required. We show on the MAPS dataset that the proposed semi-supervised CNMF method performs better than state-of-the-art low-rank factorization techniques and a little worse than supervised deep learning state-of-the-art methods, while however suffering from generalization issues.

\end{abstract}
\section{Introduction}\label{sec:introduction}

Automatic Music Transcription (AMT) is the task of transforming music recordings into symbolic format, such as scores or MIDI. It is a fundamental musical skill to acquire, taught from early age up to professional level in music schools and, given enough training, humans can be extremely accurate at transcription. Automatic music transcription aims at accelerating and improving time-consuming manual transcription and has applications in music tutoring and rehearsing, musicology analysis or in other music information retrieval tasks~\cite{benetos2018automatic}.

However, while audio generation from MIDI is rather mature, its counterpart AMT is still a very challenging task, even in scenarios involving a single multipitch instrument like a piano, which is our case study. As reported in the 2018 survey by Benetos \textit{et. al.}~\cite{benetos2018automatic}, there are mainly two families of methods to perform AMT:
1) Methods based on low-rank factorizations of spectrograms, and in particular Nonnegative Matrix Factorization (NMF). These methods are mostly unsupervised~\cite{vincent2008harmonic, cheng2016attack, gao2017polyphonic}.
2) Deep Neural Networks (DNN) which are heavily supervised. They require registered symbolic-audio training data in a large amount, which can be hard to acquire~\cite{sigtia2016end, hawthorne2017onsets, hawthorne2018enabling,kong2020high,yan2021skipping}.

A recent outbreak in the task of piano transcription (as well as other related tasks) is due to the release of the MAESTRO dataset~\cite{hawthorne2018enabling}, a large dataset of tightly matched MIDI and audio piano recordings of professional quality which improved the training quality of deep learning techniques. However, the supervised 
methods require extensive amounts of training data which may not be available for most instruments. The quality of the MAESTRO dataset comes from the existence of the Yamaha Disklavier\texttrademark, which enables co-recording of audio and MIDI. This high level technology does not exist for most instruments, and building large training dataset for most polyphonic instruments would be extremely challenging on the practical side. 

In contrast, since unsupervised factorization-based approaches do not require training data, they obviously solve the data frugality and generalization problems at the cost of being far less accurate than deep supervised approaches.

The goal of this paper is two-fold. On a first hand, leveraging training data available only in limited quantity. On another hand, deploying a variant of NMF, coined Convolutive NMF, in the context of transcription, to improve the transcription performance with respect to NMF.
The most closely related work is surely the Attack Decay model~\cite{cheng2016attack}, which also performs semi-supervision, and proposes a model reminiscent of CNMF. The major differences between the proposed CNMF framework and this work of Cheng \textit{et. al.} are discussed in Section~\ref{sec:AD}. Moreover, in Section~\ref{sec:xp}, we show that
the performance of the proposed approach are generally much higher and can reach the performance levels observed with Deep Learning at the cost of poor generalization properties. In~\cite{gao2017polyphonic}, authors also consider CNMF for piano transcription but CNMF is not the main focus of their work.

This paper is organized as follows: in Section~\ref{sec:basics}, we review the basics of NMF and CNMF for transcription. In Section~\ref{sec:learning}, semi-supervised CNMF is introduced. In Section~\ref{sec:xp} we show experimental results on MAPS and MAESTRO. Section~\ref{sec:discussion} is devoted to discussions and perspectives.

\textbf{Notations}: Matrices and higher-order arrays are denoted by capital letters, \( T_{ijk} \) is the element $(i,j,k)$ in the three-way array \( T \). To denote slices, we use semicolons, so that \( T_{i::} \) denotes for instance the slice of all elements of \( T \) on row $i$. Finally, we denote \( T_{[a:b]jk} \) elements $(i,j,k)$ with \( i\in[a,b] \).

\section{CNMF for transcription}\label{sec:basics}
\subsection{NMF and CNMF formalisms}
\begin{figure*}[ht]
  \centering
  \includegraphics[width=0.9\textwidth]{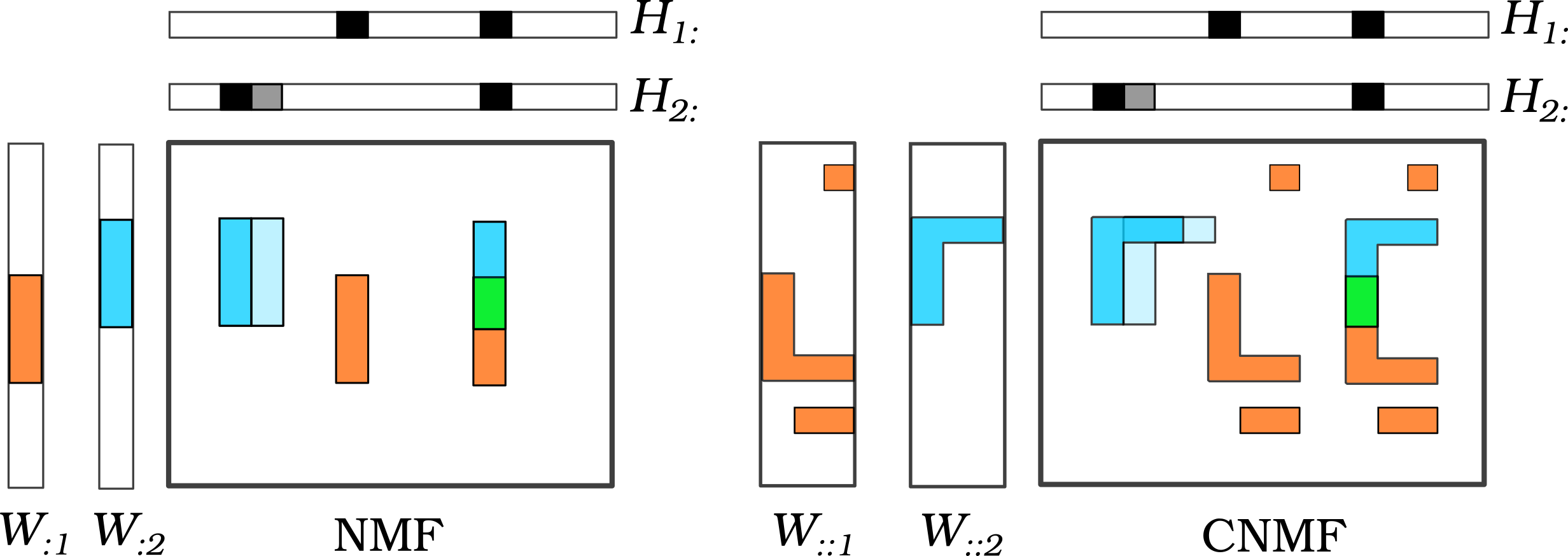}
  \caption{A visual comparison of NMF (left) and CNMF (right). CNMF allows to model complex time dependance while maintaining the number of templates low.}
  \label{fig:CNMFvsNMF}
\end{figure*}

Given an element-wise nonnegative matrix \( M \in\mathbb{R}_+^{n\times m} \) indexed as \( M_{ft} \) with \( f\in[1,n],\; t\in[1,m] \), Nonnegative Matrix Factorization (NMF) is a low-rank approximation technique that summarizes \( M \) as a sum of rank-one parts, such that
\begin{equation}
  M_{ft} = \sum_{q=1}^{r}{W_{fq}H_{qt}}
\end{equation}
where \( r\leq \min(n,m) \) is a user-defined parameter relating to the number of patterns underlying \( M \), see Figure~\ref{fig:CNMFvsNMF}.
In practice, when \( M \) is an amplitude spectrogram, such as in this work, NMF is computed approximately and boils down to solving a bi-level constrained optimization problem
\begin{equation}\label{eq:NMF}
  \underset{W\in\mathbb{R}_+^{n\times r}, H\in\mathbb{R}_+^{r\times m}}{\argmin} D_{KL}(M, WH)
\end{equation}
where \( D_{KL}(M, WH) \) is the element-wise Kullback-Leibler divergence between matrix \( M \) and its nonnegative low-rank approximation \( WH=\sum_{q=1}^{r} W_{:q}H_{q:} \).
In AMT, parameter \( r \) often relates to the number of notes expected in the recording, and therefore is generally set to (sometimes a multiple of) \( r=88 \) for piano recordings~\cite{vincent2008harmonic}.

\begin{figure}
  \hspace*{-0.5cm}\includegraphics[width=0.49\textwidth]{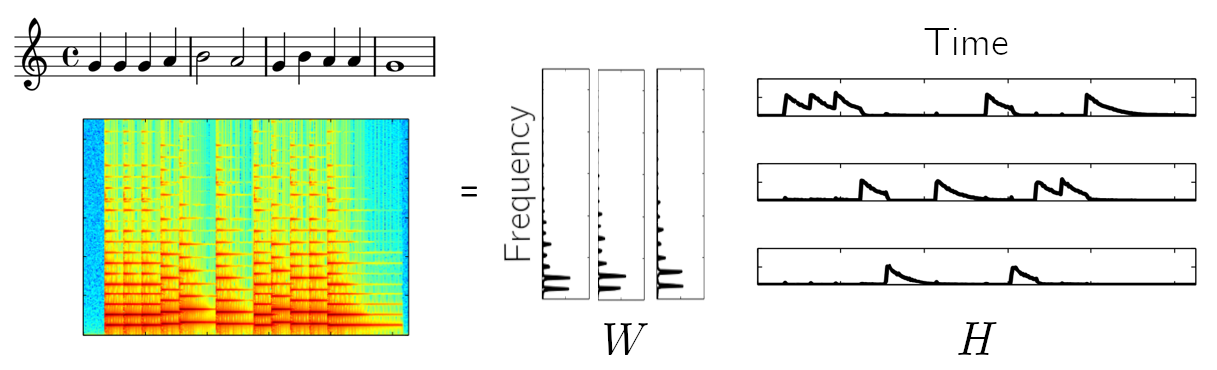}
  \caption{A toy example of transcription using NMF (adapted from~\cite{bertin2009factorisations}).}
  \label{fig:NMF}
\end{figure}

Furthermore, factor matrices \( W \) and \( H \) are respectively related to pitch and time activation.
More specifically, each column of \( W \) is expected to contain a spectral template characteristic of a single pitch on the instrument used in the recording, while each corresponding row in \( H \) is expected to provide the activation of that note in the recording~\cite{smaragdis2003non}, see Figure~\ref{fig:NMF}.

An immediate critic about applying NMF to AMT is that reducing a note to a single frequency template, even tailored for a given instrument, is too restrictive. In practice, frequency templates should evolve with both amplitude and time. While explicit amplitude dependence would break the principle of low-rank approximation underlying NMF, it is possible to extend NMF to include a time-dependence on the templates, which yields Convolutive NMF~\cite{smaragdis2006convolutive}:
\begin{equation}\label{eq:CNMF}
  \underset{W\in\mathbb{R}_+^{n\times \tau\times r}, H\in\mathbb{R}_+^{r\times m}}{\argmin} D_{KL}(M, \sum_{q=1}^{r} W_{::q} \ast H_{q:})
\end{equation}
where \(\left[W_{::q} \ast H_{q:} \right]_{:t} = \sum_{i=0}^{\tau-1} W_{:iq} H_{q(t-i)} \) is a discrete convolution and \( q\in[1,r] \), see Figure~\ref{fig:CNMFvsNMF} for an illustration. By convention, we set \(H_{q(t-i)}=0\) whenever \(t-i\leq 0\). Integer \( \tau \) is again a user-defined hyperparameter that dictates the size of the convolution window. To provide a different perspective, the element-wise noiseless CNMF also writes
\begin{equation}\label{eq:CNMFnoiseless}
  M_{ft} = \sum_{q=1}^{r}\sum_{i=0}^{\tau-1} W_{fiq}H_{q(t-i)}~.
\end{equation}

In a nutshell, CNMF enriches NMF by allowing each note to have a full STFT matrix \( W_{::q} \) as a frequency template instead of a single column. Therefore, it may also captures time-dependent events such as echoes or non-uniform partials attenuation. It can also be interpreted as a constrained NMF with large rank \( r\times \tau \) where each note is represented by \( \tau \) templates, and the corresponding \( \tau \) rows in \( H \) are constrained to be equal up to a shift. Other works have also considered enriching NMF with several templates per note albeit not using convolution, typically by fusing rows of the estimated \( H \) matrix a posteriori~\cite{wang2012score,benetos2012score,jeong2017note}.

\subsection{Comparing the Attack Decay model with CNMF}\label{sec:AD}
A reader familiar with the work of Cheng \textit{et. al.}~\cite{cheng2016attack} will notice that our work is similar in several aspects with their proposed Attack Decay (AD) framework for music transcription, but let us properly compare the models. After some rewriting of the original AD (see additional material\footnote{\url{https://github.com/cohenjer/TransSSCNMF}}), AD decomposes the data \( M \) into two terms
\begin{align}\label{eq:AD}
  M_{ft} = \sum_{q=1}^{r}\underbrace{\sum_{i=0}^{2\tau} \left(\tilde{W}^{\text{attack}}_{fq}P_{-i}\right)H_{q(t-i)}}_{\text{one note attack}} \\ + \sum_{q=1}^{r}
\underbrace{\sum_{i=\tau}^{t+\tau-1} \left(\tilde{W}^{\text{decay}}_{fq}e^{-\alpha_q(i-\tau)}\right)H_{q(t-i)}}_{\text{one note decay}}~.
\end{align}
It thus appears that the attack term is a CNMF with rank-one templates \( W^{(\text{CNMF})}_{fiq} = W^{\text{a}}_{fq}P_{i} \) which is therefore less general than the CNMF model. The decay term is also a CNMF with rank-one templates.

With some further manipulations, one can see that it is possible to entirely recast the AD model as a CNMF model with rank-two templates, which may explain the performance gap between the two models observed in Section~\ref{sec:xp}. Indeed in the semi-supervised setting, we seem to have enough data to learn unconstrained templates \( W^{\text{train}} \), and the Attack-Decay structure on the templates may not be beneficial.
\section{Template learning and CNMF}\label{sec:learning}

\subsection{Challenges in unsupervised CNMF}
In the context of music transcription, it is rarely discussed why NMF performs extremely well on simple dataset, but rather poorly on more complex ones. 
Saying that NMF, or CNMF, is a part-based representation with no destructive interferences between components
does not explain this behavior. In fact, supposing the data indeed is generated reasonably well with a ``ground-truth'' NMF \( M = AB \) for some true frequency templates \( A\in\mathbb{R}_+^{n\times r}\) and activations \(B\in\mathbb{R}_+^{r\times m} \), we need to ensure that computing an exact NMF \( M = WH \) will indeed yield \( A = W \) and \( B = H \). In other words, the data \( M \) must admit a unique NMF.

Theoretically speaking, it is known that NMF will only enjoy this uniqueness property in particular cases, such as when sources are sufficiently scattered or when the data is very sparse~\cite{Donoho2003When, fu2018identifiability}. While this may hold for simple songs where notes do not overlap a lot, in the general case one should \textbf{not} expect that \( W \) and \( H \) behave as expected without restricting the set of solutions. Even worse, CNMF being a generalized NMF model, it is bound to have even weaker uniqueness properties than NMF (but nothing is known on CNMF identifiability to the best of our knowledge). Blind CNMF has been used with additional sparsity constraints for drums transcription, but dealing with drums typically yields much sparser and lower-rank data than pitched audio due to the temporal localization of percussive sounds.

Therefore, in general, unsupervised CNMF is not regularized enough to perform transcription. While some works focus on further regularization of NMF~\cite{leplat2019minimum}, we instead turn towards semi-supervision.

\subsection{Learning note-wise templates}

Our working hypothesis is that audio recordings of isolated pitches are available, similarly to what is used for virtual instruments, except that we only make use of one template per note. Each recording is processed as the module of its complex STFT, denoted \( V_{::q}\in\mathbb{R}_+^{n\times m_q} \) where \( m_q \) is the number of STFT frames for that recording. For a regular piano one needs  \( 88 \) such templates. Apart from pitch knowledge, no registered MIDI information is required.

The goal of the learning phase here is to estimate \( W_{::q} \) for each \( q \) using each individual recording \( V_{::q} \). We propose to compute an approximate rank-one CNMF of each \( V_{::q} \) to estimate \( W_{::q} \) and \( h^{\text{train}}_q \), the latter being discarded after the training phase. From a theoretical perspective, rank-one CNMF is a constrained version of NMF of rank \( \tau \), furthermore computed on a very simple dataset. Therefore it fulfills the qualitative NMF uniqueness criteria discussed above, and we expect the recovered \( W \) to contain adequate note frequency templates.

Practically, we solve for each \( q\in[1,r] \) the following optimization problem
\begin{equation}\label{eq:CNMFrank1}
  W^{\text{train}}_{::q}, h^{\text{train}}_{q:}\in\underset{W\in\mathbb{R}_+^{n\times \tau \times 1},\; h\in\mathbb{R}_+^{1\times m}}{\argmin} D_{KL}(V_q, W\ast h)
\end{equation}
using a recently proposed multiplicative algorithm~\cite{fagot2019majorization} which alternates between \( W \) and \( h \) updates while preserving nonnegativity and ensuring cost decrease.

In spite of the rank-one approximation and the simple data, the optimization problem still proves challenging with many local minima. Therefore initialization plays an important role in the learning phase. 
Because it is reasonable to look for \(W_{::q}\) in the \( V_{::q} \) data itself, we set
\begin{equation}
  W^{\text{init}}_{::q} = V_{:[t^\ast:t^\ast+\tau-1]q}\; \text{and}\; t^\ast = \underset{t\leq m_q}{\argmax} \|V_{:[t:t+\tau-1]q}\|_1
\end{equation}
which amounts to finding the \(\tau\) consecutive columns with most energy for initialization. Then we fill \( h^{\text{init}}_{q:} \) with zeros and place a one at \(t^\ast\). Note that this initialization procedure mimics a recently proposed algorithm for separable CNMF\footnote{Separable CNMF is a computationally simpler variant of CNMF which looks for all matrices \(W_{::q}\) in the data itself.}~\cite{degleris2020provably} but is less computationally intensive. A total of 500 outer iterations are performed to learn a single note template.

\begin{figure}
  \centering
  \hspace*{-0.5cm}
  \includegraphics[width=0.49\textwidth]{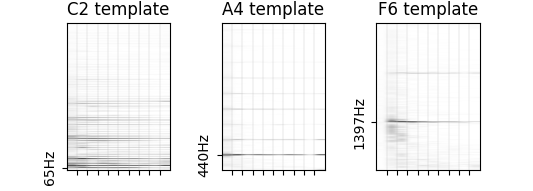}
  \caption{Three trained templates from the AkPnCGdD synthetic piano in MAPS, using \( \tau=10 \) convolution size. Templates \( W_{::q} \) have been square rooted to better highlight higher frequencies.}
  \label{fig:templates}
\end{figure}

Once the training phase is over, for a single multipitch instrument, we have at our disposal the whole dictionary \( W^{\text{train}} \), see Figure~\ref{fig:templates}.


\subsection{CNMF transcription with templates}

\begin{figure}
  \centering
  \includegraphics[width=0.45\textwidth]{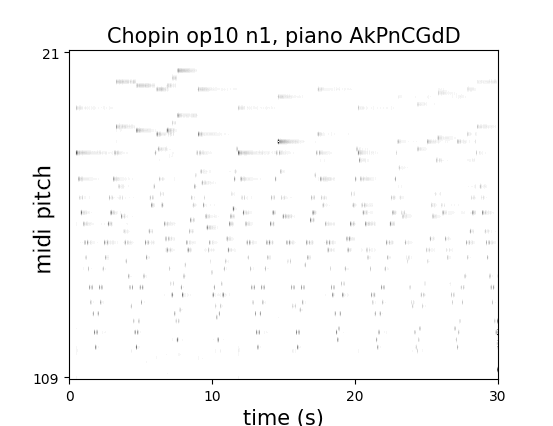}
  \caption{An example of \(H^{\text{test}}\) computed through rank-one CNMF.}
  \label{fig:H}
\end{figure}

Testing in the semi-supervised framework only consists of computing the time activations \( H \) for a given music excerpt \( M \) to transcribe, since \( W \) has been pre-trained. This makes the transcription task much easier since the problem
\begin{equation}\label{eq:CNMFtest}
  H^{\text{test}}\in\underset{H\in\mathbb{R}_+^{r\times m}}{\argmin} D_{KL}(M, \sum_{q=1}^{r} W^{\text{train}}_{::q}\ast H_{q:})
\end{equation}
is convex and therefore can be solved up to arbitrary precision with the algorithm proposed in~\cite{fagot2019majorization}. In practice 100 iterations are used, which is generally enough to reach convergence. Initialization was carried out using a few iterations of NMF with \( W \) fixed as the first column of each trained template \( W^{\text{train}}_{::q} \). An example output \( H^{\text{test}} \) is provided in Figure~\ref{fig:H}.

\subsection{Post-processing of activations}
The post-processing of \( H^{\text{test}} \) that produces a MIDI file matters  a lot. Hopefully, prior works have already proposed quite efficient post-processing using an adaptive threshold~\cite{cheng2016attack}. We essentially use the same technique but simplified.

\begin{figure}
  \centering
  \hspace*{-0.5cm}
  \includegraphics[width=0.49\textwidth]{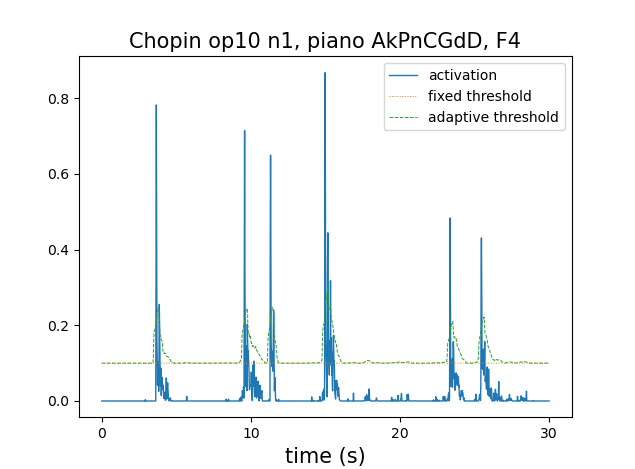}
  \caption{The CNMF activation and adaptive peak-picking method shown for note F4 using a song from MAPS.}
  \label{fig:peak_picking}
\end{figure}

In short, activation values in each row of \( H^{\text{test}} \), averaged over several consecutive frames, are added to a user-defined threshold \( \delta \), defining an adaptative threshold. An onset is detected at the position where the signal is above this adaptive threshold, see Figure~\ref{fig:peak_picking} for an illustration. Formally, an onset is detected at frame \(t\) for note \( q \) when
\begin{equation}
  h_{qt} > \frac{1}{21}\sum_{j=-10}^{10} h_{q(t+j)} + \delta,
\end{equation}
using zero-padding when necessary.
The activations are typically very sparse, so we generally did not observe spurious double peaks using the adaptive threshold contrarily to what was observed in~\cite{cheng2016attack}.

\section{Experiments on MAPS and MAESTRO}\label{sec:xp}
Although the proposed semi-supervised CNMF framework works in principle for transcribing any multipitch instrument, we only evaluate the performance for piano transcription as a proof on concept. Among the few existing open piano recordings dataset with registered audio and MIDI, in Section~\ref{sec:MAPS} we focused especially on MAPS~\cite{emiya2010maps} which has several kinds of individual notes recordings for several pianos, both virtual and acoustic. We also used MAESTRO~\cite{hawthorne2018enabling} to evaluate generalization performance in Section~\ref{sec:Maestro}. In our tests, we only considered the first 30 seconds of each song, as in~\cite{cheng2016attack}. Results are discussed in Section~\ref{sec:discussion}.

\subsection{Experimental Setup}

Let us briefly state the various experimental parameters required to reproduce the experiments\footnote{Python code to compute CNMF and reproduce all the experiments is available at \url{https://github.com/cohenjer/TransSSCNMF}}. All time signals are sampled at 44100Hz, the STFT is computed with windows of 80ms (3528 samples) with a hop-length of 20ms (882 samples). This results in \( n = 4097\) frequency bins and \( m = 1501\) time frames in the STFT for \( 30s \) of raw audio signal. No smoothing is applied to the STFT, and we set \(M\) as the amplitude spectrogram.

The \( \tau \) values are chosen among \( \tau=5, 10, 20\).
We used one template for each piano note such that \( r=88 \). Finally to fix the peak-picking threshold for \( H^{\text{test}} \), two oracle strategies are used: 1) use the same threshold for all songs, and report results for the best value on the grid \([0.01:0.01:0.4]\) 2) perform transcription with a song-dependent threshold, optimized on the same grid. The first case corresponds to a scenario where the threshold is pre-trained for a category of recording (music genre, recording conditions) while the second case corresponds to a hand-tuned threshold for a specific song to transcribe.

We compare our method with the Attack Decay (AD) model presented in Section~\ref{sec:learning} which, to the best of our knowledge, is the current state-of-the-art for unsupervised/semi-supervised piano transcription. The results reported in Table~\ref{tab:xp1} are the exact results from~\cite{cheng2016attack} (AD~\cite{cheng2016attack}), and the results of AD when applying our post-processing (AD*). In both cases transcription is performed on \( H^{\text{attack}} \) as defined in~\cite{cheng2016attack}. Despite our efforts we were unable to exactly reproduce the original AD scores. In particular the original AD paper introduces smoothness in several aspects: the data spectrograms are locally averaged, and peak-fusion is performed in the post-processing. Consequently, the AD* results enable comparison between the proposed CNMF and AD in the same pre/postprocessing conditions, while AD\cite{cheng2016attack} are the best results achieved by Cheng \textit{et. al.}. For completeness, we also report the transcription score from the state-of-the-art piano transcription network introduced in~\cite{kong2020high} which was trained on MAESTRO~\cite{hawthorne2018enabling}. 

To measure performance, we compute a notewise score using the mir\textunderscore eval~\cite{raffel2014mir_eval} toolbox with a tolerance of 50ms. The offset detection problem is not tackled. Results are shown using only F-measure (F) and Accuracy (A) metrics (reported in percent), but full results including Precision and Recalls for all pianos are available in the complementary materials online.

\subsection{Transcription performance on MAPS}\label{sec:MAPS}

The MAPS dataset contains classical piano music pieces recorded with different pianos and conditions: a Yamaha Disklavier\texttrademark in two settings 'ENSTDkCl' (EN1) and 'ENSTDkAm' (EN2), and six synthetic pianos 'AkPnBcht', 'AkPnBsdf' (AkB1-2), 'AkPnCGdD' (AkC), 'AkPnStgb' (AkS), 'SptkBGAm' (Sp), 'StbgTGd2' (St). For each piano/setting listed above, we train a template \( W^{\text{train}} \) using a rank-one CNMF as presented in Equation \ref{eq:CNMFrank1}. 
Since there are many available single notes recordings in MAPS, we chose based on performance to use the Isolated notes (ISOL / NO) recorded at Medium intensity (M).

In a first experiment, we study the sensitivity of the proposed method to the selection of the convolution window size $\tau$ and the choice of a threshold \(\delta\) tuned on the whole corpus versus on each song individually. We also compare our results to the Attack Decay model and the ByteDance supervised neural network. Results are shown in Table~\ref{tab:xp1}. From this experiment, we see that generally \(\tau=10\) performs best, and that the song-wise threshold gives better results.

In a second experiment, the templates for all other pianos are used to transcribe AkPnCGdD and ENSTDkCl to estimate the generalization capacities of the trained CNMF templates, see results in Table~\ref{tab:xp3}. Only CNMF with song-tuned threshold is shown and we set \(\tau=10\), to show the best results only.

Additionally, Table~\ref{tab:xptime} reports the average running times when training the templates and performing transcription on the AkPnCGdD recordings. This test was run on a personal computer with AMD Ryzen 5 2600\texttrademark~ processor and 16GB RAM.

\begin{table*}
  \begin{tabular}{clc|c|c|c|c|c|c|c|c|c|c|c|c|c|c|c|c|}
  \cline{4-19}
  \multicolumn{1}{l}{}                           &                                             & \multicolumn{1}{l|}{} & \multicolumn{2}{c|}{EN1}                                                  & \multicolumn{2}{c|}{EN2}                                                  & \multicolumn{2}{c|}{AkB1}                                                 & \multicolumn{2}{c|}{AkB2}                                                 & \multicolumn{2}{c|}{AkC}                                                  & \multicolumn{2}{c|}{AkS}                                                  & \multicolumn{2}{c|}{Sp}                                                   & \multicolumn{2}{c|}{St}                                                   \\ \cline{1-1} \cline{3-19}
  \multicolumn{1}{|c|}{thresh}                    & \multicolumn{1}{l|}{}                       & $\tau$                   & \cellcolor[HTML]{C0C0C0}F                       & A                       & \cellcolor[HTML]{C0C0C0}F                       & A                       & \cellcolor[HTML]{C0C0C0}F                       & A                       & \cellcolor[HTML]{C0C0C0}F                       & A                       & \cellcolor[HTML]{C0C0C0}F                       & A                       & \cellcolor[HTML]{C0C0C0}F                       & A                       & \cellcolor[HTML]{C0C0C0}F                       & A                       & \cellcolor[HTML]{C0C0C0}F                       & A                       \\ \hline
  \multicolumn{1}{|c|}{}                         & \multicolumn{1}{c|}{}                       & 5                     & \cellcolor[HTML]{C0C0C0}78                      & 65                      & \cellcolor[HTML]{C0C0C0}70                      & 55                      & \cellcolor[HTML]{C0C0C0}88                      & 80                      & \cellcolor[HTML]{C0C0C0}75                      & 62                      & \cellcolor[HTML]{C0C0C0}83                      & 72                      & \cellcolor[HTML]{C0C0C0}80                      & 69                      & \cellcolor[HTML]{C0C0C0}81                      & 70                      & \cellcolor[HTML]{C0C0C0}75                      & 61                      \\ \cline{3-19}
  \multicolumn{1}{|c|}{}                         & \multicolumn{1}{c|}{}                       & 10                    & \cellcolor[HTML]{C0C0C0}\textbf{85}             & \textbf{75}             & \cellcolor[HTML]{C0C0C0}\textbf{77}             & \textbf{64}             & \cellcolor[HTML]{C0C0C0}93                      & 88                      & \cellcolor[HTML]{C0C0C0}87                      & 78                      & \cellcolor[HTML]{C0C0C0}91                      & 84                      & \cellcolor[HTML]{C0C0C0}\textbf{88}             & \textbf{79}                      & \cellcolor[HTML]{C0C0C0}89                      & 82                      & \cellcolor[HTML]{C0C0C0}84                      & 74                      \\ \cline{3-19}
  \multicolumn{1}{|c|}{}                         & \multicolumn{1}{c|}{\multirow{-3}{*}{CNMF}} & 20                    & \cellcolor[HTML]{C0C0C0}83                      & 72                      & \cellcolor[HTML]{C0C0C0}76                      & 63                      & \cellcolor[HTML]{C0C0C0}\textbf{94}             & \textbf{89}             & \cellcolor[HTML]{C0C0C0}\textbf{87}             & \textbf{79}             & \cellcolor[HTML]{C0C0C0}\textbf{92}             & \textbf{86}             & \cellcolor[HTML]{C0C0C0}87                      & 79                      & \cellcolor[HTML]{C0C0C0}\textbf{90}             & \textbf{83}             & \cellcolor[HTML]{C0C0C0}\textbf{86}             & \textbf{77}             \\ \cline{2-19}
  \multicolumn{1}{|c|}{\multirow{-4}{*}{global}} & \multicolumn{2}{c|}{AD*}                                            & \multicolumn{1}{l|}{\cellcolor[HTML]{C0C0C0}81} & \multicolumn{1}{l|}{69} & \multicolumn{1}{l|}{\cellcolor[HTML]{C0C0C0}68} & \multicolumn{1}{l|}{53} & \multicolumn{1}{l|}{\cellcolor[HTML]{C0C0C0}66} & \multicolumn{1}{l|}{50} & \multicolumn{1}{l|}{\cellcolor[HTML]{C0C0C0}71} & \multicolumn{1}{l|}{56} & \multicolumn{1}{l|}{\cellcolor[HTML]{C0C0C0}60} & \multicolumn{1}{l|}{43} & \multicolumn{1}{l|}{\cellcolor[HTML]{C0C0C0}67} & \multicolumn{1}{l|}{51} & \multicolumn{1}{l|}{\cellcolor[HTML]{C0C0C0}64} & \multicolumn{1}{l|}{47} & \multicolumn{1}{l|}{\cellcolor[HTML]{C0C0C0}67} & \multicolumn{1}{l|}{50} \\ \hline
  \multicolumn{1}{|c|}{}                         & \multicolumn{1}{l|}{}                       & 5                     & \cellcolor[HTML]{C0C0C0}82                      & 70                      & \cellcolor[HTML]{C0C0C0}74                      & 59                      & \cellcolor[HTML]{C0C0C0}90                      & 82                      & \cellcolor[HTML]{C0C0C0}80                      & 69                      & \cellcolor[HTML]{C0C0C0}87                      & 78                      & \cellcolor[HTML]{C0C0C0}84                      & 74                      & \cellcolor[HTML]{C0C0C0}86                      & 77                      & \cellcolor[HTML]{C0C0C0}81                      & 69                      \\ \cline{3-19}
  \multicolumn{1}{|c|}{}                         & \multicolumn{1}{l|}{}                       & 10                    & \cellcolor[HTML]{C0C0C0}\textbf{88}             & \textbf{79}             & \cellcolor[HTML]{C0C0C0}\textbf{80}             & \textbf{68}             & \cellcolor[HTML]{C0C0C0}\textbf{95}             & \textbf{91}             & \cellcolor[HTML]{C0C0C0}\textbf{90}             & \textbf{83}             & \cellcolor[HTML]{C0C0C0}94                      & 89                      & \cellcolor[HTML]{C0C0C0}\textbf{90}             & \textbf{82}             & \cellcolor[HTML]{C0C0C0}\textbf{93}             & \textbf{87}             & \cellcolor[HTML]{C0C0C0}89                      & 80                      \\ \cline{3-19}
  \multicolumn{1}{|c|}{}                         & \multicolumn{1}{l|}{\multirow{-3}{*}{CNMF}} & 20                    & \cellcolor[HTML]{C0C0C0}85                      & 75                      & \cellcolor[HTML]{C0C0C0}78                      & 66                      & \cellcolor[HTML]{C0C0C0}\textbf{95}             & \textbf{91}             & \cellcolor[HTML]{C0C0C0}\textbf{90}             & \textbf{83}             & \cellcolor[HTML]{C0C0C0}\textbf{94}             & \textbf{90}                      & \cellcolor[HTML]{C0C0C0}89                      & 81                      & \cellcolor[HTML]{C0C0C0}92                      & 87                      & \cellcolor[HTML]{C0C0C0}\textbf{89}             & \textbf{81}             \\ \cline{2-19}
  \multicolumn{1}{|c|}{\multirow{-4}{*}{song}}   & \multicolumn{2}{c|}{AD*}                                            & \multicolumn{1}{l|}{\cellcolor[HTML]{C0C0C0}82} & \multicolumn{1}{l|}{70} & \multicolumn{1}{l|}{\cellcolor[HTML]{C0C0C0}69} & \multicolumn{1}{l|}{54} & \multicolumn{1}{l|}{\cellcolor[HTML]{C0C0C0}68} & \multicolumn{1}{l|}{52} & \multicolumn{1}{l|}{\cellcolor[HTML]{C0C0C0}73} & \multicolumn{1}{l|}{59} & \multicolumn{1}{l|}{\cellcolor[HTML]{C0C0C0}61} & \multicolumn{1}{l|}{45} & \multicolumn{1}{l|}{\cellcolor[HTML]{C0C0C0}69} & \multicolumn{1}{l|}{54} & \multicolumn{1}{l|}{\cellcolor[HTML]{C0C0C0}66} & \multicolumn{1}{l|}{50} & \multicolumn{1}{l|}{\cellcolor[HTML]{C0C0C0}70} & \multicolumn{1}{l|}{54} \\ \hline
  \multicolumn{3}{|c|}{AD~\cite{cheng2016attack} {}{}}                                                                                      & \cellcolor[HTML]{C0C0C0}82                      & 70                      & \cellcolor[HTML]{C0C0C0}-                       & -                       & \cellcolor[HTML]{C0C0C0}-                       & -                       & \cellcolor[HTML]{C0C0C0}-                       & -                       & \cellcolor[HTML]{C0C0C0}85                      & 74                      & \cellcolor[HTML]{C0C0C0}-                       & -                       & \cellcolor[HTML]{C0C0C0}-                       & -                       & \cellcolor[HTML]{C0C0C0}-                       & -                       \\ \hline
  \multicolumn{3}{|c|}{ByteDance DNN\cite{kong2020high} {}{}}                                                                                     & \multicolumn{1}{l|}{\cellcolor[HTML]{C0C0C0}\textbf{89}} & \multicolumn{1}{l|}{\textbf{81}} &
  \multicolumn{1}{l|}{\cellcolor[HTML]{C0C0C0}77} & \multicolumn{1}{l|}{65} &
  \multicolumn{1}{l|}{\cellcolor[HTML]{C0C0C0}\textbf{98}} & \multicolumn{1}{l|}{\textbf{97}} &
  \multicolumn{1}{l|}{\cellcolor[HTML]{C0C0C0}\textbf{95}} & \multicolumn{1}{l|}{\textbf{90}} &
  \multicolumn{1}{l|}{\cellcolor[HTML]{C0C0C0}\textbf{98}} & \multicolumn{1}{l|}{\textbf{96}} &
  \multicolumn{1}{l|}{\cellcolor[HTML]{C0C0C0}87} & \multicolumn{1}{l|}{77} &
  \multicolumn{1}{l|}{\cellcolor[HTML]{C0C0C0}\textbf{97}} & \multicolumn{1}{l|}{\textbf{95}} &
  \multicolumn{1}{l|}{\cellcolor[HTML]{C0C0C0}\textbf{95}} & \multicolumn{1}{l|}{\textbf{90}} \\ \hline
  \end{tabular}
  \caption{CNMF, AD, and the ByteDance supervised network performance with respect to the choice of hyperparameter \(\tau\) and the choice of the peak-picking threshold, without training/testing mismatch for CNMF and AD. Only the first 30s of each songs were used. AD* uses the same pre/post-processing as CNMF. Tolerance is 50ms.}
  \label{tab:xp1}
\end{table*}

\begin{table}
  \begin{tabular}{cc|c|c|c|c|c|c|}
    \cline{3-8}
      &                           & EN2                        & AkB1                       & AkB2                       & AkS                        & Sp                         & St                         \\ \hline
    \multicolumn{1}{|c|}{}                      & \cellcolor[HTML]{C0C0C0}F & \cellcolor[HTML]{C0C0C0}74 & \cellcolor[HTML]{C0C0C0}77 & \cellcolor[HTML]{C0C0C0}77 & \cellcolor[HTML]{C0C0C0}70 & \cellcolor[HTML]{C0C0C0}74 & \cellcolor[HTML]{C0C0C0}77 \\ \cline{2-8}
    \multicolumn{1}{|c|}{\multirow{-2}{*}{AkC}} & A                         & 59                         & 64                         & 63                         & 56                         & 59                         & 63                         \\ \hline
    \multicolumn{1}{|c|}{}                      & \cellcolor[HTML]{C0C0C0}F & \cellcolor[HTML]{C0C0C0}76 & \cellcolor[HTML]{C0C0C0}67 & \cellcolor[HTML]{C0C0C0}68 & \cellcolor[HTML]{C0C0C0}69 & \cellcolor[HTML]{C0C0C0}67 & \cellcolor[HTML]{C0C0C0}69 \\ \cline{2-8}
    \multicolumn{1}{|c|}{\multirow{-2}{*}{EN1}} & A                         & 62                         & 50                         & 52                         & 53                         & 52                         & 53                         \\ \hline
  \end{tabular}
  \caption{Transcription scores for CNMF, with training/testing mismatch.}
  \label{tab:xp3}
\end{table}

\begin{table}[]
  \centering
\begin{tabular}{l|c|l|l|c|}
\cline{2-5}
    & \multicolumn{3}{c|}{Training}                                                & \multicolumn{1}{l|}{Transcription} \\ \hline
\multicolumn{1}{|c|}{$\tau$} & 5                        & \multicolumn{1}{c|}{10} & \multicolumn{1}{c|}{20} & 10                                 \\ \hline
\multicolumn{1}{|l|}{Av. time}            & \multicolumn{1}{l|}{56s} & 193s                    & 634s                    & 239s                               \\ \hline
\end{tabular}
\caption{Average computation time for a learning pattern (Training) or transcribing 30s of a song (Transcription) for semi-supervised CNMF. Results are reported for the AkPnCGdD piano in MAPS.}
\label{tab:xptime}
\end{table}

\subsection{Generalization on MAESTRO}\label{sec:Maestro}
A natural question regarding CNMF templates is how well they can be used outside their training context without any domain adaptation. While results shown in Table~\ref{tab:xp3} already provide a partial answer, we also tried to apply CNMF to the MAESTRO dataset. However, since no individual notes recordings are publicly available for MAESTRO, we used the templates learnt from MAPS. We transcribed 20 songs from the MAESTRO test set randomly chosen.

The results are quite poor: even when choosing song-wise thresholds, for all templates, CNMF does not reach above 59\% in F-measure (test results are available in the supplementary materials). For comparison, the state-of-the-art with supervised deep learning techniques reaches above 95\% F-measure on MAESTRO. Its performance on MAPS with data augmentation are also state-of-the-art, around 89\% F-measure on EN1, despite the training/testing mismatch.

\section{Discussion}\label{sec:discussion}
In light of the experiments conducted in Section~\ref{sec:xp}, let us discuss the strengths of the proposed CNMF. It exhibits a significant improvement with respect to the Attack Decay model, which as far as we know is state-of-the-art for semi-supervised piano transcription. This is even more true when using the same pre-processing and post-processing for AD and CNMF, the former being in particular prone to unstable activations which were not observed in the latter.
We may therefore affirm that the improvement in performance is indeed due to the CNMF model design. In other words, CNMF with a semi-supervised setting is an efficient piano transcription method. From numerical results, it seems that a convolution window size \(\tau=10\) is a good compromise between quality of transcription and transcription computation time.

The CNMF method does not perform better than the supervised state-of-the-art method we denoted as ByteDance DNN, which is expected given that this neural-network competitor is trained on MAESTRO which contains more than two hundred hours of perfectly aligned MIDI and audio piano recording of professional level. We still reach similar performances on some pianos such as EN1, EN2 and AkS. Nevertheless, the ByteDance DNN is not trained on MAPS contrarily to the proposed semi-supervised CNMF.

Moreover, the proposed semi-supervised setting only requires a handful of training dataset which are relatively easy to acquire. Indeed, only individual notes recordings are necessary, without any audio and MIDI registration. Compared to the very large amount of data currently required by state-of-the-art deep learning approaches, this is a huge advantage of the proposed approach applicable to any acoustic instrument with well-defined onsets readily available. Sadly our study is limited to piano transcription. A perspective of this work is to apply it to transcribe polyphonic instruments for which recording registered MIDI and audio is challenging.

Finally, while the performance does depend on the choice of a good activation threshold, CNMF still performs well using a global threshold over all songs in MAPS for each piano. Therefore extensively tuning the threshold hyperparameter is not essential to the success of CNMF here.

Despite these encouraging results, CNMF has a few issues which open interesting perspectives. First, it clearly has a significant generalization problem, or in other words, the learning stage overfits the training data. From Table~\ref{tab:xp3}, it appears that a mismatch between training and testing inside MAPS, while detrimental to transcription performance, is not as severe as a learning on MAPS and testing on MAESTRO. A tentative explanation is that the MAESTRO recordings are live performances with quite loud reverberation, while the MAPS recordings are drier. Looking for an audio transformation of the templates that minimizes recording conditions mismatch would therefore probably prove beneficial to generalize pre-recorded CNMF templates. Retraining a template library given few annotated data in the testing set could also be a possible solution. Whether this domain adaptation can be done fully blindly is still unclear however.

Second, despite performance not relying too much on the threshold level, the threshold selection method on the other hand is extremely important. Using a fixed threshold instead of the adaptive peak-picking drastically decreased performances in our early tests. But this also means that the post-processing of activations can be further improved using more involved technique than thresholding each note individually, and this research direction should not be overlooked if transcription performances of CNMF are to be further improved.

Third, for simplicity only one template for each note was used for the transcription phase. However, most instruments sound quite differently depending on how they are played. The proposed semi-supervised framework currently does not account for this timbre variation with amplitude or technique, and adapting the current method to make use of several templates per notes is an interesting research direction.

Finally according to the results shown in Table~\ref{tab:xptime}, computation time is rather large even in the testing phase. With the current implementation, real-time processing is therefore prohibited. Using a CNMF solver dedicated to Kullback-Leibler divergence or working on a more efficient rank-one CNMF solver than~\cite{fagot2019majorization} could nevertheless drastically reduce computation time.

\section{Conclusion}
The state-of-the-art for automatic piano transcription is undeniably nowadays detained by deep learning techniques. However these methods rely on very large audio and symbolic registered dataset which are potentially very hard to obtain. In this work, we propose a competitive semi-supervised matrix factorization model which only requires labeled recordings of each individual notes. We show that when there is no mismatch between the training data and the test data, our approach performs significantly better than semi-supervised state-of-the-art approaches, approaching supervised deep learning performance. Therefore, we believe that using CNMF instead of NMF is an important step towards learning more reasonable frequency templates in low-rank approximation techniques for piano transcription or other similar tasks. Further works should however be devoted to adapt pre-trained templates to reduce generalization error. Improving the onset detection method, allowing timbre variation in templates and reducing computation time are other important research directions. Finally, the proposed semi-supervised approach should be tested with other instruments than the piano and in a multi-instrument setup.

\begin{acknowledgments}
Jeremy E. Cohen and Axel Marmoret thank ANR JCJC LoRAiA ANR-20-CE23-0010 for supporting this work.

\end{acknowledgments}

\bibliography{all_refs}

\end{document}